\documentclass[a4paper,pra,twocolumn,showpacs,showkeys,amsmath,amssymb,aps]{revtex4-1}

\usepackage{dsfont}
\usepackage{hyperref}
\usepackage{alltt}

\newtheorem{definition}{Definition}
\newtheorem{theorem}{Theorem}

\newtheorem{example}{Example}
\newtheorem{proposition}{Proposition}

\newcommand{\eg}{\emph{e.g.\/}}
\newcommand{\ie}{\emph{i.e.\/}}
\newcommand{\etal}{\emph{et al.}}
\newcommand{\Mathematica}{\emph{Mathematica}}
\newcommand{\Matlab}{\emph{Matlab}}
\newcommand{\Octave}{\emph{GNU Octave}}

\newcommand{\SWAP}{\ensuremath{\mathrm{SWAP}}}
\newcommand{\M}{\ensuremath{\mathbb{M}}}
\DeclareMathOperator{\tr}{tr}

\DeclareMathOperator{\wek}{\mathbf{vec}}
\DeclareMathOperator{\res}{\mathbf{res}}

\renewcommand{\Vec}[1]{\ensuremath{\mathbf{#1}}}
\newcommand{\Mn}[1]{\ensuremath{\M_n({#1})}}

\newcommand{\C}{\ensuremath{\mathds{C}}}
\newcommand{\Cplx}{\ensuremath{\C}}

\newcommand{\ket}[1]{\ensuremath{|#1\rangle}}
\newcommand{\bra}[1]{\ensuremath{\langle#1|}}
\newcommand{\ketbra}[2]{\ensuremath{\ket{#1}\bra{#2}}}

\newcommand{\1}{{\rm 1\hspace{-0.9mm}l}}
\newcommand{\Id}{\1}

\newcommand{\halmos}{\newline\vspace{3mm}\hfill $\Box$}
\providecommand{\proof}{\noindent {\it Proof.\ }}

\newcommand{\scalar}[2]{\left( #1 , #2 \right)}

\newcommand{\qi}{\texttt{QI}}

\begin{document}
\title{Singular value decomposition and matrix reorderings in quantum
information theory}

\author{Jaros{\l}aw Adam Miszczak}
\affiliation{Institute of Theoretical and Applied Informatics, Polish Academy
of Sciences, Ba{\l}tycka 5, 44-100 Gliwice, Poland}

\begin{abstract}
We review Schmidt and Kraus decompositions in the form of singular value
decomposition using operations of reshaping, vectorization and reshuffling. We
use the introduced notation to analyse the correspondence between quantum states
and operations with the help of Jamio{\l}kowski isomorphism. The presented matrix
reorderings allow us to obtain simple formulae for the composition of quantum
channels and partial operations used in quantum information theory. To provide
examples of the discussed operations we utilize a package for the \Mathematica\
computing system implementing basic functions used in the calculations related
to quantum information theory.

\end{abstract}

\date{08/04/2011 (v.~1.30)}

\maketitle

\section{Introduction}
Quantum information theory~\cite{ingarden76qit,NC00,qiphysics} aims to provide
methods of harnessing the quantum nature of information carriers to develop more
efficient quantum algorithms and more secure communication protocols.
Mathematically quantum systems are described using the formalism of density
matrices and the most general form of quantum evolution is described by
completely positive operators~\cite{NC00,BZ06}.

In many situations in quantum information theory one deals with vector spaces of
the tensor-product form. For example, the description of composite quantum
systems is based on a tensor product of spaces describing sub-systems. This is
essential for the phenomenon of quantum entanglement, which is one of the most
important features of quantum information theory~\cite{horodecki09entanglement,
guhne09detection}. Also the theory of quantum channels, which are used, for
example, to describe errors in quantum computation and communication protocols,
deals with the composite channels that are described as tensor products of
channels. The composition of quantum channels gives rise to another phenomenon
unique to quantum information theory, namely the non-additivity of channel
capacity~\cite{hastings09counterexample,brandao09hastings}.

The main aim of this report is to present a uniform view on Schmidt and 
Kraus decompositions. Both decompositions provide very important tools used in
quantum information theory. Schmidt decomposition is used to describe quantum
entanglement in the special case of pure quantum states. Kraus decomposition, on
the other hand, is used in the analysis of quantum channels. We achieve our goal
by presenting both decompositions in the form of singular value decomposition and
by using some matrix reorderings. As the reorderings are used in many
branches of mathematics, physics and computer science, our goal is also to
clarify the used notation. The presented concepts form the basis for the
package of functions for \Mathematica\ computer algebra systems presented in
the last part of this report.

This report is organized as follows. In Section \ref{sec:algebra} we review some
basic algebraic facts applied in the this report. In particular we fix the notions
of matrix reshaping, vectorization and reshuffling. In Section
\ref{sec:schmidt-vec} we use the SVD theorem in the finite dimensional Hilbert
space to obtain Schmidt decomposition for pure quantum states. We also
rephrase Schmidt decomposition in any unitary space and apply it to density
matrices to obtain so called \emph{operator Schmidt decomposition}. In
Section~\ref{sec:channels} we use the conditions for quantum channels and Singular
Value Decomposition to derive the Kraus form of a quantum channel and we analyse
the composition of quantum channels and partial operations. Finally,
Appendix~\ref{app:mathematica} contains some examples of the discussed concepts
using the package of functions for \Mathematica\ computing system.

\paragraph*{Notation}
In what follows we denote by $\Vec{v}$ elements of finite vector space and
$\ket{\phi}$ pure states. By $\M_{m,n}$ we denote the set of all $m\times n$
matrices over $\Cplx$. The set of square $n\times n$ matrices is denoted by
$\M_n$. The set of $n$-dimensional density matrices (normalized, positive
semi-definite operators on~$\Cplx^n$) is denoted by~$\Omega_n$.

The set $\M_{n}$ has the structure of the Hilbert space with the scalar product
given by the formula
\begin{equation}
\scalar{A}{B} = \tr A^\dagger B.
\end{equation}
This particular Hilbert space is known as the Hilbert-Schmidt space of operators 
acting on $\Cplx^n$ and we will denote it by $\mathcal{H}_\mathrm{HS}$.

\section{Singular value decomposition and matrix reorderings}\label{sec:algebra}
In this section we review some basic algebraic facts used in the following parts
of this report. One should note that the operations of reshaping and
vectorization, introduced in this section, are used in many areas of science and
engineering -- see for example \cite{HPS,vanloan00ubiquitous}. For this reason
in many cases the naming conventions differ depending on the authors' preferences
and backgrounds.

\subsection{Singular value decomposition}
For the sake of consistency we start by recalling singular value decomposition 
(SVD) which is valid for any $n\times m$ matrix over 
$\Cplx$~\cite[Chapter 7.3]{hj1}.

\begin{theorem}[Singular Value Decomposition]
Let $A\in \M_{m,n}$ has the rank $k\leq m$. Then there exist unitary matrices 
$U\in \M_m$ and $V\in \M_n$ such that
\begin{equation}\label{eqn:svd}
A=U\Sigma V^\dagger.
\end{equation}
The matrix $\Sigma = \{\sigma_{ij}\} \in \M_{m,n}$ is such that
\begin{equation}
\sigma_{ij}=0,\ \mathrm{for}\ i\not=j,
\end{equation}
and
\begin{equation}
\sigma_{11}\geq\sigma_{22}\geq \ldots \geq\sigma_{kk}>\sigma_{k+1,k+1}
= \ldots = \sigma_{qq} = 0,
\end{equation}
with $q=\min(m,n)$. 
\end{theorem}

The numbers $\sigma_{ii}\equiv \sigma_{i}$ are \emph{singular vales}, \ie\
non-negative square roots of the eigenvalues of $AA^\dagger$. The columns of $U$
are eigenvectors of $AA^\dagger$ and the columns of $V$ are eigenvectors of
$A^\dagger A$.

In the special case when $A$ is positive semi-definite the above decomposition
is equivalent to the eigendecomposition of~$A$.

\subsection{Reshaping and vectorization}
Singular value decomposition provides us with the particular form of a given
matrix. This can be directly applied in the case when we deal with linear maps
on a finite-dimensional vector space.

In order to use the singular value decomposition we need one more algebraic tool, 
namely the mapping between $\M_{m,n}$ and $\M_{mn,1}$ (or $\Cplx^{mn}$). We 
define two functions, which can be used as such mappings.

\begin{definition}[Reshaping and vectorization]
Let $A=[a_{ij}]_{i,j}\in \M_{m,n}(\Cplx)$. We define the \emph{reshaping} of 
$A$ as
\begin{eqnarray}
\res(A) &=& (a_{11},a_{12},\ldots,a_{1n},a_{21},a_{22},\ldots\\\nonumber
&&\ldots,a_{2n},\ldots,a_{m1},a_{m2},\ldots,a_{mn})^T,
\end{eqnarray}
and the \emph{vectorization} of $A$ as
\begin{eqnarray}
\wek(A) &=& (a_{11},a_{21},\ldots,a_{m1},a_{12},a_{22},\ldots\\\nonumber
&&\ldots,a_{m2},\ldots,a_{1n},a_{2n},\ldots,a_{mn})^T
\end{eqnarray}
where `${\ }^T$' denotes matrix transposition.
\end{definition}

In other words the vectorization of matrix $A$ means its reordering in column order
and reshaping -- in row order. Note that $\res(A)$ is equivalent to $\wek(A^T)$. 
Both operations can be achieved using, for example, \texttt{Flatten}
function in \Mathematica\ or \texttt{reshape} function in \Matlab\ and \Octave.   

Both $\wek(A)$ and $\res(A)$, map $\M_{m,n}$ onto $\M_{mn,1}$. Both
operations can also be interchanged as they are connected by the formula
\begin{equation}
\res{A}=\wek{A^T}.
\end{equation}
Thus, it is rather a matter of taste which one to use.

One should keep in mind that there are
several notational conventions in literature for denoting vectorization and
reshaping operations. We use the definition of vectorization as provided in
\cite{vanloan00ubiquitous}, \cite[Definition 7.1.1]{bernstein05matrix}
and~\cite[Definition 4.2.9]{hj2}. The reshaping operation defined above agrees
with the convention used in~\cite[Chapter 10]{BZ06} and it corresponds to the
row-major order method of turning a matrix into a vector. In \cite{Havel2003}
this operation is denoted as $\mathbf{col}$. In the recent preprint of
Gilchrist \etal\ \cite{Gilchrist2009} the authors refer to $\res$ operation
defined above as to vectorization. 

Vectorization and reshaping have many useful properties, some of which we are 
going to use in the following sections. In particular if $A,B,C\in \M_{m}$ then we 
have the following.
{
\renewcommand{\labelenumi}{(P\arabic{enumi})}
\begin{enumerate}
  \item $\wek(A) = \res(A^T)$ for $A\in \M_m$,\label{prop:vec-res-interchange}
  \item $\wek(ABC) = (C^T\otimes A) \wek(B)$ and $\res(ABC) = (A\otimes C^T)
    \res(B)$,\label{prop:telegraphic}
  \item $\wek(AB) = (\Id\otimes A)\wek(B) = (B^T\otimes
    \Id)\wek(A)$,\label{prop:telegraphic-id}
  \item $\wek(A\circ B) = \wek(A)\circ\wek(B)$, where `$\circ$' denotes the
    Hadamard product \cite{hj1},\label{prop:hadamard}
  \item $\tr A^\dagger B = \wek(A)^*\cdot\wek(B) = \res(A)^*\cdot\res(B)$, where
    `$\cdot$' denotes the scalar product of two vectors in $\C^n$.
\end{enumerate}
}

In particular properties (P\ref{prop:telegraphic}),
(P\ref{prop:telegraphic-id}) and (P\ref{prop:hadamard}) from the above list also hold
for rectangular matrices of appropriate dimensions. According to \cite{hj2} the
property (P\ref{prop:telegraphic}) appeared for the first time in
\cite{roth34direct} and it will be crucial in the next sections.

\subsection{Reshuffling}
Our main goal is to use SVD to analyse composite quantum states and the 
dynamics of quantum systems. In both cases we need to deal with the 
tensor-product structure. For pure quantum states this structure is fixed by 
the physical structure of the system we aim to describe. For quantum channels 
this structure is introduced by Jamio{\l}kowski isomorphism, which uses
the operation of \emph{reshuffling}.

Reshuffling can be used to fix particular tensor product structure in the set 
of matrices. Roughly speaking a reshuffled matrix is a matrix represented in a 
particular tensor-product base. 

Let us denote by $\{\epsilon_i:i=1,\ldots,m^2\}$ and $\{\varepsilon_j:j=1,
\ldots,n^2\}$ canonical bases in $\M_m$ and $\M_n$ respectively. This is to say
that $\res(\epsilon_i)$ ($\res(\varepsilon_j)$) has 1 at $i$-th ($j$-th)
position and zeros elsewhere and $\res(\epsilon_{i}\otimes \varepsilon_{j})$
has $1$ at the $i\times j$-th position.

\begin{definition}[Reshuffling]\label{def:reshuffle}
Let $A\in\M_{k}$ with $k=mn$, \ie\ $\M_{k}=\M_m\otimes\M_n$. Matrix with
elements
\begin{equation}
\{A^{R(m,n)}\}_{ij}\stackrel{\mathrm{def}}{=}
\tr\left[ (\epsilon_i\otimes\varepsilon_j)^\dagger A\right]
\end{equation}
is called a \emph{reshuffling of matrix $A$ with respect to subspaces $\M_m$
and $\M_n$}.
\end{definition}

Using $\res$ operation a reshuffled matrix can be expressed as
\begin{equation}
\{A^{R(m,n)}\}_{ij}= \res\left( \epsilon_i\otimes\varepsilon_j\right) \cdot \res(A),
\end{equation}
where we have used the fact that matrices $\epsilon_i$ and $\varepsilon_j$ are
real.

Note that this type of matrix reordering was introduced without any connection 
to quantum physics in~\cite{oxenrider85reorderings}.

One can also introduce the reshuffling operation using transposed canonical
bases, which are ordered accordingly in column order. This is to say that
$\wek(\epsilon^T_i)$ ($\wek(\varepsilon^T_j)$) has 1 at $i$-th ($j$-th) position
and zeros elsewhere and $\wek(\epsilon^T_{i}\otimes \varepsilon^T_{j})$ has $1$
at the ($i\times j$)-th position. We define alternative reshuffling by counting
matrix elements in column order.

\begin{definition}[Alternative reshuffling]\label{def:alt-reshuffle}
Let $A\in\M_{k}$ with $k=mn$, \ie\ $\M_{k}=\M_m\otimes\M_n$. Matrix with 
elements
\begin{equation}
\{A^{R'(m,n)}\}_{ij}\stackrel{\mathrm{def}}{=}
\tr\left[ (\epsilon_j^T\otimes\varepsilon_i^T)^\dagger A\right]
\end{equation}
is called an \emph{alternative reshuffling of matrix $A$ with respect to 
subspaces $\M_m$ and $\M_n$}.
\end{definition}

Taking into account that base matrices are real we get
\begin{equation}
\{A^{R'(m,n)}\}_{ij}=\tr\left[ (\epsilon_j\otimes\varepsilon_i) A\right].
\end{equation}

Note that it is also possible to define reshuffling in more general case when
matrix $A$ is of the form $A=X\otimes Y$ with $X\in\M_{k,l}$ and $Y\in\M_{m,n}$. 

We usually work with a density matrix $\rho$, which is said to be 
an element of $S(\mathcal{H}_A\otimes\mathcal{H}_B)$. In such case the 
reshuffling operation is understood with respect to canonical bases in 
$S(\mathcal{H}_A)$ and $S(\mathcal{H}_B)$ 

Moreover, while working with $\rho\in\M_n\otimes\M_n$ we write simply $\rho^R$ as
long as the dimensions of matrices in question can be deduced from the context.

\begin{example}
To give a simple example of reshuffling operation one can use a square matrix
$A\in\M_{n^2}$. For example, if $A\in\M_4$ it is given as
\begin{equation}
A=\left(
\begin{array}{cccc}
 \alpha_{1,1} & \alpha_{1,2} & \alpha_{1,3} & \alpha_{1,4} \\
 \alpha_{2,1} & \alpha_{2,2} & \alpha_{2,3} & \alpha_{2,4} \\
 \alpha_{3,1} & \alpha_{3,2} & \alpha_{3,3} & \alpha_{3,4} \\
 \alpha_{4,1} & \alpha_{4,2} & \alpha_{4,3} & \alpha_{4,4}
\end{array}
\right),
\end{equation}
then we have
\begin{equation}
A^{R(2,2)}=\left(
\begin{array}{cccc}
 \alpha_{1,1} & \alpha_{1,2} & \alpha_{2,1} & \alpha_{2,2} \\
 \alpha_{1,3} & \alpha_{1,4} & \alpha_{2,3} & \alpha_{2,4} \\
 \alpha_{3,1} & \alpha_{3,2} & \alpha_{4,1} & \alpha_{4,2} \\
 \alpha_{3,3} & \alpha_{3,4} & \alpha_{4,3} & \alpha_{4,4}
\end{array}
\right).
\end{equation}
On the other hand taking the alternative definition of reshuffling we get
\begin{equation}
A^{R'(2,2)}=\left(
\begin{array}{cccc}
 \alpha_{1,1} & \alpha_{3,1} & \alpha_{1,3} & \alpha_{3,3} \\
 \alpha_{2,1} & \alpha_{4,1} & \alpha_{2,3} & \alpha_{4,3} \\
 \alpha_{1,2} & \alpha_{3,2} & \alpha_{1,4} & \alpha_{3,4} \\
 \alpha_{2,2} & \alpha_{4,2} & \alpha_{2,4} & \alpha_{4,4}
\end{array}
\right).
\end{equation}
\end{example}

\begin{example}
Reshuffling operation is a linear map on $\M_n$ and as such it can be
represented as a matrix. For example the operation $\rho\mapsto{\rho}^{R(2,2)}$
on $\M_4$ has the following matrix representation 
\begin{equation}\label{eqn:reshuffle-matrix-form}
M_{R(2,2)}=
\left(
\begin{smallmatrix}
 1 & 0 & 0 & 0 & 0 & 0 & 0 & 0 & 0 & 0 & 0 & 0 & 0 & 0 & 0 & 0
   \\
 0 & 1 & 0 & 0 & 0 & 0 & 0 & 0 & 0 & 0 & 0 & 0 & 0 & 0 & 0 & 0
   \\
 0 & 0 & 0 & 0 & 1 & 0 & 0 & 0 & 0 & 0 & 0 & 0 & 0 & 0 & 0 & 0
   \\
 0 & 0 & 0 & 0 & 0 & 1 & 0 & 0 & 0 & 0 & 0 & 0 & 0 & 0 & 0 & 0
   \\
 0 & 0 & 1 & 0 & 0 & 0 & 0 & 0 & 0 & 0 & 0 & 0 & 0 & 0 & 0 & 0
   \\
 0 & 0 & 0 & 1 & 0 & 0 & 0 & 0 & 0 & 0 & 0 & 0 & 0 & 0 & 0 & 0
   \\
 0 & 0 & 0 & 0 & 0 & 0 & 1 & 0 & 0 & 0 & 0 & 0 & 0 & 0 & 0 & 0
   \\
 0 & 0 & 0 & 0 & 0 & 0 & 0 & 1 & 0 & 0 & 0 & 0 & 0 & 0 & 0 & 0
   \\
 0 & 0 & 0 & 0 & 0 & 0 & 0 & 0 & 1 & 0 & 0 & 0 & 0 & 0 & 0 & 0
   \\
 0 & 0 & 0 & 0 & 0 & 0 & 0 & 0 & 0 & 1 & 0 & 0 & 0 & 0 & 0 & 0
   \\
 0 & 0 & 0 & 0 & 0 & 0 & 0 & 0 & 0 & 0 & 0 & 0 & 1 & 0 & 0 & 0
   \\
 0 & 0 & 0 & 0 & 0 & 0 & 0 & 0 & 0 & 0 & 0 & 0 & 0 & 1 & 0 & 0
   \\
 0 & 0 & 0 & 0 & 0 & 0 & 0 & 0 & 0 & 0 & 1 & 0 & 0 & 0 & 0 & 0
   \\
 0 & 0 & 0 & 0 & 0 & 0 & 0 & 0 & 0 & 0 & 0 & 1 & 0 & 0 & 0 & 0
   \\
 0 & 0 & 0 & 0 & 0 & 0 & 0 & 0 & 0 & 0 & 0 & 0 & 0 & 0 & 1 & 0
   \\
 0 & 0 & 0 & 0 & 0 & 0 & 0 & 0 & 0 & 0 & 0 & 0 & 0 & 0 & 0 & 1
\end{smallmatrix}
\right).
\end{equation}
\end{example}

One can note that reshuffling and alternative reshuffling are 
connected by the relation~\cite[Chapter 10]{BZ06}
\begin{proposition}
For any $A\in\M_m\otimes\M_n$ we have
\begin{equation}
A^{R'} = (SA^{R}S)^T,
\end{equation}
where $S$ is the swap operation.
\end{proposition}

In the next section we use the following simple fact connecting reshuffling
and the tensor product.
\begin{proposition}\label{fact:kron-reshuffle}
Let $A\in\M_n$, $B\in\M_m$. Then we have
\begin{equation}\label{eqn:kron-reshuffle}
\res\big( (A\otimes B)^R \big)= \res (A)\otimes \res (B)
\end{equation}
and
\begin{equation}\label{eqn:kron-reshuffle-prim}
\wek\big( (A\otimes B)^{R'} \big)= \wek (A)\otimes \wek (B).
\end{equation}
\end{proposition}

Proposition~\ref{fact:kron-reshuffle} follows directly from the definition of
reshuffling and it allows us to interchange between product base in
Hilbert-Schmidt space and the one in $\Cplx^n$.

\section{Schmidt decomposition} \label{sec:schmidt-vec}
Now we are ready to use the introduced tools for deriving some important results
from quantum information theory.

Our first goal is to prove a particular representation of vectors in finite 
dimensional vector space with inner product. This representation is known in 
quantum information theory as \emph{Schmidt decomposition}~\cite{BZ06, ziman}.

Schmidt decomposition was first stated for an infinite-dimensional Hilbert
space~\cite{schmidt07zur, pietsch10erhard}, but it is more often used in a
version which deals with finite-dimensional spaces only. It is frequently used
in quantum information theory to distinguish between separable and entangled
states~\cite{guhne09detection}.

\subsection{Schmidt decomposition for pure states}\label{sec:schmidt-pure}
We start with Schmidt decomposition for pure states, \ie\ unit vectors in
a finite-dimensional Hilbert space $\Cplx^{mn}=\Cplx^{m}\otimes\Cplx^{n}$. This
form is used in quantum information theory to study quantum entanglement.

\begin{theorem}\label{theo:schmidt-kets}
Any pure state $\ket{\psi}\in \Cplx^m \otimes \Cplx^n$ can be represented as
\begin{equation}\label{eqn:schmidt-form-kets}
\ket{\psi} = \sum_{i=1}^{k} \sqrt{\lambda_i}\ket{\alpha_i}\otimes \ket{\beta_i},
\end{equation}
where $\{\ket{\alpha_i}\}\in \Cplx^m$ and $\{\ket{\beta_i}\}\in \Cplx^n$ are 
orthogonal in respective Hilbert spaces and $k\leq\min(m,n)$.
\end{theorem}

\proof
We can always represent $\ket{\psi}\in \Cplx^m \otimes \Cplx^n$ using canonical 
basis as
\begin{equation}
\ket{\psi} = \sum_{i=1}^m \sum_{j=1}^n C_{ij} \ket{e_i}\otimes \ket{f_j},
\end{equation}
where $\{\ket{e_i}\}\in\Cplx^m$ and $\{\ket{f_j}\}\in\Cplx^m$ are canonical 
bases in respective subspaces, $C\in \M_{m,n}$ and vectors 
$\ket{e_i}\otimes\ket{f_j},\ i=1,\ldots,m,\ j=1,\ldots,n$ have the following form
\begin{equation}
\ket{e_i}\otimes\ket{f_j} = (0,\ldots,0,1,0,\ldots,0)^T,
\end{equation}
with $1$ at position $ij$ and zeros elsewhere. In this particular basis 
$\ket{\psi} = \res(C)$. Using the SVD and the property (P\ref{prop:telegraphic}) for the reshaping
operation we get
\begin{equation}
\ket{\psi}= \res(U\Sigma V^\dagger) = (U\otimes V^*) \res(\sigma_{ij}\delta_{ij})
\end{equation}
and by using the canonical basis we get 
\begin{eqnarray}
\ket{\psi }
&=& (U\otimes V^*) \sum_{i=1}^m \sum_{j=1}^n \sigma_{ij}\delta_{ij} \ket{b_{ij}}\\
&=& \sum_{i=1}^k \sigma_{ii} U\ket{e_{i}}\otimes V^*\ket{f_{i}},
\end{eqnarray}
where $k$ is the order of $C$. Since $\sigma_{ii}$ are square roots of 
eigenvalues of positive matrix $CC^\dagger$ we can write
\begin{equation}
\ket{\psi} = \sum_{i=1}^k \sqrt{\lambda_i}\ket{\alpha_{i}}\otimes \ket{\beta_{i}},
\end{equation}
with $\sigma_{ii}=\sqrt{\lambda_i}$, $\ket{\alpha_{i}} = U\ket{e_i}$ and 
$\ket{\beta_{i}} = V^*\ket{f_i}$. 
\halmos

\begin{definition}
Number $k$ of elements in Schmidt decomposition is often referred to as 
the \emph{Schmidt number}. 
\end{definition}

States of bipartite systems are among the most interesting objects in quantum
information theory. This is because the tensor-product structure of state space
results in the presence of states which cannot be mimicked using classical
theory. These special states are called \emph{entangled states} and are used in
quantum protocols and algorithms.

Theorem \ref{theo:schmidt-kets} allows us to distinguish entangled pure states
from non-entangled (or separable) pure states. We have the following theorem,
which provides us with the simplest \emph{separability
criterion}~\cite{horodecki09entanglement, guhne09detection}.

\begin{theorem}
Pure state is separable iff its Schmidt number is equal to 1.
\end{theorem}

\subsection{Schmidt decomposition for unitary spaces}\label{sec:schmidt-general}
As one can easily see the line of reasoning used in the proof of Theorem
\ref{theo:schmidt-kets} can be repeated for any finite-dimensional vector space
$\mathcal{H}$ with scalar product. All we need is a particular representation of
elements in this space in the base of the tensor-product form. This is to say
that $\mathcal{H}$ has to be of the form $\mathcal{H} =\mathcal{H}_A \otimes
\mathcal{H}_B$. Moreover, we do not need Hilbert spaces to spell-out this
theorem. We require only for $\mathcal{H}_A$ and $\mathcal{H}_B$ to be finite-dimensional
vectors spaces over $\C$ with inner product, \ie\ $\mathcal{H}_A$
and $\mathcal{H}_B$ have to be unitary spaces. Thus we can easily reformulate
Schmidt decomposition in somehow more universal language. 

\begin{theorem}[Schmidt decomposition]\label{theo:schmidt}
Let $\mathcal{H}_A$ and $\mathcal{H}_B$ be unitary spaces. Any element $\Vec{v}
\in \mathcal{H}=\mathcal{H}_A \otimes \mathcal{H}_B$ can be represented as
\begin{equation}
\Vec{v} = \sum_{i=1}^{k} \sqrt{\lambda_i}\Vec{u}_i\otimes \Vec{w}_i,
\label{eqn:schmidt-form-general}
\end{equation}
where vectors $\Vec{u}_i\in \mathcal{H}_A$ and $\Vec{w}_i\in\mathcal{H}_B$,
$i=1,2,\ldots,k$ are mutually orthogonal in respective spaces and
$k\leq\min(\dim\mathcal{H}_A,\dim \mathcal{H}_B)$.  
\end{theorem}

\proof The line of reasoning is analogous to the one used to prove 
Theorem~\ref{theo:schmidt-kets}. In this case $\lambda_i$ are singular values of 
the matrix 
\begin{equation}
C_{ij}=(\Vec{e}_i\otimes \Vec{e}_j,\Vec{v}).
\end{equation}
with $(\cdot,\cdot)$ being scalar product on $\mathcal{H} = \mathcal{H}_A
\otimes \mathcal{H}_B$.
\halmos

This form of Schmidt decomposition allows us to use it not only for pure 
states, but also for any space with an introduced scalar product. In many 
situations it is convenient, however, to use the isomorphism defined by 
reshaping (or vectorization). 

Recently Schmidt decomposition applied to two-qubit unitary gates was used
to study properties of this particular set~\cite{balakrishnan10operator}. Using
this tool it was found, for example, that locally equivalent non-local gates
posses the same set of Schmidt coefficients.

\subsection{Example: bipartite density matrices}
In quantum mechanics only a small fraction of states can be represented by 
normalized vectors in some Hilbert space $\Cplx^n$. Especially when we are 
interested in interactions of the system in question with the environment, we have 
to represent states of the system as density matrices, \ie\ positive operators
with unit trace.

As an example of Theorem~\ref{theo:schmidt} we will use
Theorem~\ref{theo:schmidt} to analyse the space of bipartite density matrices.
Let us start by recalling the definition of bipartite separable
state~\cite{werner89epr,guhne09detection} 
\begin{definition}
Let $\rho$ be a state $\rho$ of bipartite quantum system described
$\rho\in\mathcal{S}(\mathcal{H}_A\otimes\mathcal{H}_B)$. We say that $\rho$ is
\emph{separable} if it can be represented as a convex combination 
\begin{equation}
\rho=\sum_{i=1}^k p_i \rho_i^{(A)}\otimes\rho_i^{(B)},
\end{equation}
with $\sum p_i=1$, and for all $i=1,\ldots,k$ we have
$\rho_i^{(A)}\in\mathcal{S}(\mathcal{H}_A)$ and
$\rho_i^{(B)}\in\mathcal{S}(\mathcal{H}_B)$. If $\rho$ cannot be represented in
this form we say that it is \emph{entangled}. 
\end{definition}

Although the space $\Omega_m\otimes \Omega_n$ of density matrices 
in not a~vector space, we can exploit the linear structure it inherits as a~subset 
of $\mathcal{H}_\mathrm{HS} = \M_{mn}$.

Any element $\rho\in\Omega_m\otimes\Omega_n\subset \M_{mn}$ can be 
written using the standard basis as
\begin{equation}
\rho = \sum_{i=1}^{m^2}\sum_{k=1}^{n^2} C_{ij} \epsilon_{i}\otimes \varepsilon_{j},
\end{equation}
where $\epsilon_i\in\M_m, i=1,\ldots,m^2$ and $\varepsilon_j\in\M_n, 
j=1,\ldots,n^2$ are standard bases in respective spaces. 

Using Schmidt decomposition \ref{theo:schmidt} we can rewrite $\rho$ as
\begin{equation}\label{eqn:operator-schmidt}
\rho = \sum_{l=1}^{k} \sigma_l \epsilon'_{l}\otimes \varepsilon'_{l},
\end{equation}
where $\sigma_i$ are singular values of the matrix 
\begin{equation}
C_{ij} =\tr\left[ \rho^\dagger (\epsilon_{i}\otimes \varepsilon_{j}) \right]=
 \res(\rho)^*\cdot \res(\epsilon_{i}\otimes \varepsilon_{j}).
\end{equation}

This representation can be also obtained using the isomorphism 
$\M_{mn}\simeq\Cplx^{m^2n^2}$ and represent elements of $\M_n$ as vectors in 
$\Cplx^{m^2n^2}$ with the help of reshaping operation
\begin{equation}
\ket{X} \stackrel{\mathrm{def}}{=}\res X,
\end{equation}
so that $\ket{X}\in\Cplx^{m^2n^2}$ for $X\in\M_{mn}$. Using this reasoning one 
can see directly how to construct base vectors in 
Eq.~(\ref{eqn:operator-schmidt})~\cite[Lemma 10.1]{BZ06}.

The representation given by Eq.~(\ref{eqn:operator-schmidt}) is sometimes referred to
as an \emph{operator Schmidt decomposition}~\cite{nielsenPhD}, but as one can
see it provides only an example of the application of Schmidt decomposition as
presented in Theorem~\ref{theo:schmidt}.


\section{Quantum channels}\label{sec:channels}
Since we aim to apply singular value decomposition to quantum channels, we need
to introduce some basic facts about them. We restrict ourselves to the
finite-dimensional case and the special subclass of trace preserving (TP)
quantum channels.

A state in quantum mechanics is described using density matrices and thus any
quantum evolution $\Phi$ has to transform initial density matrix
$\rho_\mathrm{in}\in\Omega_m$ into density matrix
$\Phi(\rho_\mathrm{in})=\rho_\mathrm{out}\in\Omega_n$.

\subsection{Definitions}
The set of quantum operations has some particular structure~\cite{BZ06,ziman}.
First of all we assume that any such map $\Phi:\rho_\mathrm{in} \mapsto
\rho_\mathrm{out}$ has to be linear. This is motivated by that fact that any
mixed state can be represented as a convex combination of other states in
infinitely many possible ways. The linearity of quantum channel $\Phi$ means that its
action does not depend on the particular representation of input density matrix. 

The main condition, however, for a linear map to be a proper quantum operation 
follows from the positivity of input and output states. In order to get more 
information about the form of $\Phi$ we need to use some physical arguments. 

It is clear that any physical map $\Phi$ (\ie\ any operation that 
can be implemented in a laboratory) has to preserve positivity. However, by 
performing an operation $\Phi$ on our system, we perform $\Phi\otimes \Id$ on 
our system and on environment. As such, any physical map has to be 
\emph{completely positive} (CP), \ie\ any extension of $\Phi$ of the form 
$\Phi\otimes\Id_m$ with $\Id_m\in\M_m$ and $m=1,2,\ldots$ has to be positive.

\begin{definition}[CP-map]
A map $\Phi$ is called completely positive (CP) if it preserves positivity and
for any $n=1,2,\ldots$ the map
\begin{equation}
\Phi\otimes\Id_n,
\end{equation}
where $\Id_n$ is an identity operation on $n$-dimensional space of states, also 
preserves positivity.
\end{definition}

This definition introduces the Kronecker product of channels, which is described
in more detailes in Sec.~\ref{sec:channels-composition}.

Using the above definition we can define \emph{quantum channel}, which describes 
the most general form of the evolution of quantum systems.

\begin{definition}[Quantum channel]
Any CP-map preserving trace is called a \emph{quantum channel} or a \emph{quantum 
operation}.
\end{definition}


We use the isomorphism $\M_{mn}\simeq\Cplx^{m^2n^2}$ to calculate the elements
of its matrix representation. As a linear map any $\Phi : \Omega_n \mapsto\Omega_n$, 
$\Phi(\rho_\mathrm{in})=\rho_\mathrm{out}$, can be written as a~matrix 
$M_\Phi\in \M_{n^2}$
\begin{equation}\label{eqn:matrix-operation}
\res(\rho_\mathrm{out}) = M_\Phi \res(\rho_\mathrm{in}),
\end{equation}
where
\begin{eqnarray}
M_\Phi &=& \left\{(\epsilon_k,\Phi(\epsilon_l))\right\}_{k,l=1,\ldots,n^2}\\\nonumber
&=& \left\{\tr \left[\epsilon_k^\dagger \Phi(\epsilon_l)\right]\right\}_{k,l=1,\ldots,n^2}.
\end{eqnarray}
has $n^4$ elements. Here again we have used canonical basis 
$\{\epsilon_k\}_{k=1,\ldots,n^2}$ in $\M_{n^2}$.

Surprisingly more information about the positivity of a given map can be
obtained if we represent map $\Phi$ in a specific basis, namely the one obtained
as a~tensor product of base matrices in subspaces of dimension $n^2$. To exploit
this structure we define so called \emph{dynamical matrix} of the map $\Phi$.
\begin{definition}[Dynamical matrix]
Let $\Phi$ be a linear map on $\M_n$. The dynamical matrix for $\Phi$ is defined
as a matrix $D_\Phi\in\M_{n^2}$
\begin{equation}
D_\Phi = \left\{\tr\left[ (\epsilon_{i}\otimes \epsilon_{j}) M_\Phi\right]
\right\}_{i,j=1,\ldots,n},
\end{equation}
where $\{\epsilon_{i}\}_{i=1,\ldots,n}$ is a canonical basis in $\M_n$, or, 
equivalently
\begin{equation}
D_\Phi=M_\Phi^{R(n,n)}.
\end{equation}
\end{definition}

One should note that the elements of the matrix $D_\Phi$ can be calculated
according to the formula
\begin{equation}\label{eqn:map-indeces-reshuffle}
\bra{(i-1) n  + j }D_\Phi\ket{(k-1) n + l} 
= \tr \left[(\epsilon_{i}\otimes \epsilon_{j})^\dagger \Phi(\epsilon_{k}\otimes
\epsilon_{l})\right].
\end{equation}

Note that this allows to use a four-index notation as introduced in \cite{BZ06}.
This notation allows to express the idea behind reshuffling as (see \cite[Eqn.
11.25]{BZ06})
\begin{equation}
\bra{k} \Phi(\ketbra{i}{j})\ket{l} = \bra{k \otimes i} D_{\Phi}\ket{l \otimes j} .
\end{equation}

For quantum information theory the most important fact expressed using the 
dynamical matrix is known as \emph{Choi theorem}.

\begin{theorem}[Choi \cite{choi75completely}]\label{teo:choi}
Linear map $\Phi$ is completely positive iff $D_\Phi$ is positive.
\end{theorem}

This theorem allows us to check easily if a given map is completely positive. The
detailed discussion of the CP conditions is presented in~\cite{ziman,BZ06} for
one-qubit quantum channels and in \cite{chencinska09cp-qutrits} for one-qutrit
channels.

\begin{example}
The operation of matrix transposition $\mathrm{T}(\rho)=\rho^T$ on 
$\Cplx^2$ can be expressed as
\begin{equation}\label{eqn:mtx-transpose}
M_\mathrm{T}=
\left(
\begin{smallmatrix}
 1 & 0 & 0 & 0 \\
 0 & 0 & 1 & 0 \\
 0 & 1 & 0 & 0 \\
 0 & 0 & 0 & 1
\end{smallmatrix}
\right),
\end{equation}
which is equivalent to \SWAP\ for a two-qubit system. In this case we have
$D_\mathrm{T} = M_\mathrm{T}^R = M_\mathrm{T}$ and, since the spectrum of this
matrix is $\{-1,1,1,1\}$, we can see that the transposition is not completely
positive.
\end{example}

In general, the transposition operation can be introduced on $\M_{m,n}$. The
general form of this operation is given by the following
theorem~\cite[Th.~4.3.8]{hj2}.

\begin{theorem}
For a matrix $A \in \M_{mn}$ there exists a unique matrix $P(m,n)$ such that
\begin{equation}
\res A^T = P(m,n)\res A
\end{equation}
given by formula
\begin{equation}
P(m,n) = \sum_{i=1}^m\sum_{j=1}^n \epsilon_{ij}^{\phantom{T}}\otimes\epsilon_{ij}^T
\end{equation}
where $\{\epsilon_{ij}\}$, with $i=1,\ldots,m$ and $j=1,\ldots,n$ is a standard
basis in $\M_{mn}$ and $P(m,n)$ is a permutation matrix.
\end{theorem}

Note that, as it preserves the spectrum, the transposition is a positive map.
Operations which are positive, but not completely positive play, an important
role in quantum information theory since they are used to detect quantum
entanglement~\cite{guhne09detection}.

Another interesting feature of quantum theory is the correspondence between quantum
states and quantum channels~\cite{zyczkowski04duality}.

The dynamical matrix for the operations $\Phi$ is defined as $D_\Phi=M_\Phi^R$,
where `${}^R$' denotes a \emph{reshuffling} operation~\cite{BZ06}. The dynamical
matrix for the trace-preserving operation acting on $N$-dimensional system is
an~$N^2\times N^2$ positive defined matrix with trace~$N$. We can introduce
the natural correspondence between such matrices and density matrices on $N^2$ by
normalizing $D_\Phi$. Such a correspondence is known as \emph{Jamio{\l}kowski
isomorphism}~\cite{jamiolkowski72linear, zyczkowski04duality}.

Let $\Phi$ be a~completely positive trace-preserving map acting on density
matrices. We define Jamio\l{}kowski matrix of $\Phi$ as
\begin{equation} \label{def:jam}
 \rho_\Phi = \frac{1}{N}D_\Phi. 
\end{equation}

Jamio\l{}kowski matrix has the same mathematical properties as a quantum state
\ie{} it is a semi-definite positive matrix with a trace equal to one. It is
sometimes referred to as \emph{Jamio\l{}kowski operation
matrix}~\cite{jamiolkowski72linear}.

\subsection{Kraus decomposition}
Now we are ready to use the singular value decomposition to obtain a special
representation of quantum channels known as the \emph{Kraus form}. As we will
see Kraus decomposition of a operations is obtained as Schmidt 
decomposition of its linear representation.

Let us now consider quantum channel $\Phi$ acting on $\Omega_n$. Its matrix 
representation $M_\Phi$ is an element of $\M_{n^2}$ and so is its dynamical
matrix $D_\Phi$.

One can represent $D_\Phi$ in the basis
\begin{equation}
\{\epsilon_i\otimes\epsilon_j:i,j=1,\ldots,n^2\},
\end{equation}
composed of tensor products of elements of canonical bases in $\M_n$. We get
\begin{equation}
M_\Phi = \sum_{i=1}^{n^2} \sum_{j=1}^{n^2} D_{\Phi_{ij}} 
\epsilon_i\otimes\epsilon_j.
\end{equation}

By taking into account the fact that the matrix $D_\Phi$ is positive and by using 
Schmidt decomposition (Theorem~\ref{theo:schmidt}) we get
\begin{equation}
M_\Phi = \sum_{i=1}^{k}{\sigma_i} \kappa_i\otimes\kappa_i^*,
\end{equation}
where $\kappa_i,\ i=1,\ldots,k$ are mutually orthogonal elements of
Hilbert-Schmidt space of operators. Recall that $M_\Phi$ acts on
$\rho\in\Omega_n$ according to Eq.~\ref{eqn:matrix-operation}. Combining this
with Property~\ref{prop:telegraphic} we get
\begin{eqnarray}
\res(\rho_\mathrm{out})&=&\res\Phi(\rho_\mathrm{in})= M_\Phi\res(\rho_\mathrm{in})\\\nonumber 
&=& \sum_{i=1}^{k}\sigma_i \kappa_i\otimes\kappa_i^* \res(\rho_\mathrm{in})\\\nonumber
&=&\sum_{i=1}^k \sigma_i \res (\kappa_i \rho_\mathrm{in}\kappa_i^\dagger)\\\nonumber
&=& \res \left(\sum_{i=1}^k \sigma_i\kappa_i \rho_\mathrm{in}\kappa_i^\dagger\right).
\end{eqnarray}

Thus we have obtained the following representation of quantum channels.
\begin{theorem}[Kraus form]
Any CP map $\Phi:\Omega_N\rightarrow\Omega_N$ can be represented as
\begin{equation}
\Phi(\rho) = \sum_{i=1}^k \sigma_i\kappa_i\rho \kappa_i^\dagger,
\end{equation}
where $\kappa_i$ are un-reshaped singular vectors of $D_\Phi$ and $\sigma_i$ are 
singular values of $D_\Phi$.
\end{theorem}

For an alternative proof based on Stinespring dilatation theorem see
\eg~\cite{ziman}.

Operators $\{K_i=\sqrt{\sigma_i}\kappa_i:i=1,2,\ldots,k\}$ in the above 
decomposition are known as \emph{Kraus operators}.

The Kraus form of a quantum channel is non-unique. We can choose another set
of operators $\{\nu_i:i=1,\ldots,l\}$ such that it represents an action of
channel $\Phi$, \ie\
\begin{equation}
\Phi(\rho_{\mathrm{in}})  = \sum_{i=1}^l\nu_i \rho_\mathrm{in}\nu_i^\dagger.
\end{equation}
Operators $K_i=\sqrt{\sigma_i}\kappa_i$ are usually referred to as canonical 
Kraus operators.

\begin{example}\label{ex:depolar-kraus}
Let us consider the completely depolarizing channel
$\Delta_{n,p}:\Omega_n\mapsto\Omega_n$~\cite{hayashi} defined as
\begin{equation}
\Delta_{n,p}(\rho) = p \rho +(1-p)  \frac{\Id}{n} \tr \rho,
\end{equation}
with $n=1,2,\ldots$ and $0\leq p \leq 1$. Depolarizing channel acting on initial
state
\begin{equation}
\rho_\mathrm{in} = 
\left(
\begin{smallmatrix}
a & b+ic\\
b-ic & 1-a
\end{smallmatrix}
\right)
\end{equation}
results in an output state
\begin{equation}
\rho_\mathrm{out} = 
\left(
\begin{smallmatrix}
 \frac{1}{2}+\left(a-\frac{1}{2}\right) p & (b+i c) p \\
 (b-i c) p & \frac{1}{2} -(a-\frac{1}{2})p
\end{smallmatrix}
\right).
\end{equation}

In one-qubit case the dynamical matrix of $\Delta_{2,p}$ reads
\begin{equation}
D_{\Delta_{2,p}} = 
\left(
\begin{smallmatrix}
 \frac{p+1}{2} & 0 & 0 & p \\
 0 & \frac{1-p}{2} & 0 & 0 \\
 0 & 0 & \frac{1-p}{2} & 0 \\
 p & 0 & 0 & \frac{p+1}{2}
\end{smallmatrix}
\right),
\end{equation}
and it has singular values
\begin{equation}
\left\{\frac{p}{2},\frac{p}{2},\frac{p}{2},\frac{1}{2} (4-3 p)\right\}.
\end{equation}
Un-reshaped singular vectors of $D_{\Delta_{2,p}}$ are 
\begin{equation}
\left\{\left(
\begin{smallmatrix}
 -\frac{1}{\sqrt{2}} & 0 \\
 0 & \frac{1}{\sqrt{2}}
\end{smallmatrix}
\right),\left(
\begin{smallmatrix}
 0 & 1 \\
 0 & 0
\end{smallmatrix}
\right),\left(
\begin{smallmatrix}
 0 & 0 \\
 1 & 0
\end{smallmatrix}
\right), \left(
\begin{smallmatrix}
 \frac{1}{\sqrt{2}} & 0 \\
 0 & \frac{1}{\sqrt{2}}
\end{smallmatrix}
\right)\right\},
\end{equation}
and we obtain the following collection of Kraus operators
\begin{equation}
\left\{
\left(
\begin{smallmatrix}
 -\frac{\sqrt{p}}{2} & 0 \\
 0 & \frac{\sqrt{p}}{2}
\end{smallmatrix}
\right),\left(
\begin{smallmatrix}
 0 & \sqrt{\frac{p}{2}} \\
 0 & 0
\end{smallmatrix}
\right),\left(
\begin{smallmatrix}
 0 & 0 \\
 \sqrt{\frac{p}{2}} & 0
\end{smallmatrix}
\right),\left(
\begin{smallmatrix}
  \frac{1}{2}\sqrt{4- 3 p} & 0 \\
 0 & \frac{1}{2}\sqrt{4- 3 p}
\end{smallmatrix}
\right)\right\}.
\end{equation}

It can be also easily checked that $\Delta_{2,p}$ can be also represented by
Kraus operators~\cite{NC00}
\begin{equation}
\left\{
\frac{\sqrt{1+3p}}{2}\Id,
\frac{\sqrt{1-p}}{2}\sigma_x,
\frac{\sqrt{1-p}}{2}\sigma_y,
\frac{\sqrt{1-p}}{2}\sigma_z
\right\},
\end{equation}
where $\sigma_x,\sigma_y$ and $\sigma_z$ are Pauli matrices. This representation
is more appealing from the physical point of view.
\end{example}

Using Kraus representation we can characterize specific types of quantum
channels. First of all we can distinguish a class of trace-preserving
operations.

\begin{definition}[Trace-preserving map]
A channel $\Psi$ given as a collection of Kraus operators $\{A_i\}_{i=1}^n$ is
trace-preserving if
\begin{equation}
\sum_{i=1}^n A_i^{\phantom{\dagger}} A_i^\dagger = \Id.
\end{equation}
\end{definition}

Another important class of quantum channels are random unitary channels.

\begin{definition}[Random unitary map]
A channel $\Phi$ is called a \emph{random unitary} if it can be represented as
\begin{equation}
\Phi(\rho) = \sum_{i=1}^k p_i U_i \rho U_i^\dagger,
\end{equation}
where operators $U_i, i=1,\ldots,k$ are unitary, $0\leq p_i, i=1,\ldots.k$ and $\sum_ip_i=1$.
\end{definition}

An important example of a random unitary channel is given by \emph{generalized
Pauli channel}~\cite{hayashi}, which is an extension to any dimension of the
one-qubit Pauli channel. 

\begin{example}[Generalized Pauli channel]
We define two families of unitary operators:
\begin{equation}
X_d=\sum_0^{d-1} \ket{j-1\bmod d}\bra{j},
\end{equation} 
and 
\begin{equation}
Z_d=\mathrm{diag}\left(1,e^{2i \pi/d \times 1},\ldots e^{2i \pi/d \times (d-1)}\right).
\end{equation} 
The action of generlized Pauli channel  $\Pi_d$ of dimension $d$ is defined as
\begin{equation}
\Pi_d(\rho)=\sum_{i,j=0}^{d-1} p_{i,j} X_d^i Z_d^j \rho (X_d^i Z_d^j)^\dagger,
\end{equation}
where $0\leq p_{i,j}\leq 1$ and $\sum p_{i,j}=1$.
\end{example}

Generalized Pauli channel is an example of unital channel, \ie\ it satisfies the
condition $\Pi_d(\Id)=\Id$.

\subsection{Composition of channels}\label{sec:channels-composition}
To this point we have been dealing with simple quantum channels (\ie\
channels acting on the whole analysed system) only. However, some features unique to
quantum information theory can be observed when one deals with \emph{composite}
quantum channels.

Choi theorem~\ref{teo:choi} deals with the extensions of a given map to a higher
dimensional space. Such extensions are maps on $\M_m\otimes\M_n$. 

\begin{definition}[Composite channel]\label{def:composite-channel}
Let $\Phi$ and $\Psi$ be quantum channels. Quantum channel $\Phi\otimes\Psi$ is
defined using its matrix representation as
\begin{equation}\label{eqn:superoper-composite}
M_{\Phi\otimes\Psi} = M_{R^{-1}}(M_\Phi \otimes M_\Psi)M_R,
\end{equation}
where $M_{R^{-1}}=M_R^{-1}=M_R$ is the matrix representation of the reshuffling
map $\rho\mapsto\rho^R$ given in~Eq.~(\ref{eqn:reshuffle-matrix-form}) or,
equivalently, as a channel acting on the initial state $\rho$ as
\begin{equation}
\res\left((\Phi\otimes\Psi)(\rho)\right)=
\left(M_\Phi\otimes M_\Psi\left(\res\left(\rho^R\right)\right)\right)^R.
\end{equation}
\end{definition}

As the reshuffling operation represents the exchange between canonical and
tensor-product base, this definition is simply the standard definition known 
from the standard multi-linear algebra.

\subsection{Partial operations}
Representation (\ref{eqn:superoper-composite}) can be used to calculate 
composition $\Phi\otimes\Psi$ of any two quantum channels $\Phi$ and $\Psi$.
If we take one of them to be identity $\Psi=\Id$ we get so-called \emph{partial
operations}.

\begin{definition}[Partial operation]\label{def:partial-operation}
Let $\Phi$ be a quantum channel acting on $m$-dimensional state space. We say
that the channel
\begin{equation}
\Phi\otimes\Id_n
\end{equation}
is a partial application of $\Phi$ on  $m\times n$ dimensional space or that it 
is an extension of $\Phi$ to $m\times n$ dimensional space.
\end{definition}

Partial operations are used extensively in quantum information theory,
especially in the context of quantum
entanglement~\cite{horodecki09entanglement}.

Let us return to the transposition operation on one-qubit system and let us see 
how it behaves under the extension to a two-qubit system. 
\begin{example}[Partial transposition]
We define a partial transposition on the first subsystem as $\mathrm{T}_1 =
\mathrm{T}\otimes\Id_2$. This map has a matrix representation
\begin{equation}\label{eqn:ptrace-def}
M_{\mathrm{T}_1}=M_\mathrm{R^{-1}}(M_\mathrm{T}\otimes\Id_4) M_\mathrm{R},
\end{equation}
where $M_\mathrm{R}=M_\mathrm{R^{-1}}$ is a matrix representation of the
reshuffling operations given by Eq.~(\ref{eqn:reshuffle-matrix-form}) and matrix
representation of transposition $M_\mathrm{T}$ is given by
Eq.~(\ref{eqn:mtx-transpose}). In this case
\begin{equation}\label{eqn:ptrace-matrix}
M_{\mathrm{T}_1}=
\left(
\begin{smallmatrix}
 1 & 0 & 0 & 0 & 0 & 0 & 0 & 0 & 0 & 0 & 0 & 0 & 0 & 0 & 0 & 0
   \\
 0 & 1 & 0 & 0 & 0 & 0 & 0 & 0 & 0 & 0 & 0 & 0 & 0 & 0 & 0 & 0
   \\
 0 & 0 & 0 & 0 & 0 & 0 & 0 & 0 & 1 & 0 & 0 & 0 & 0 & 0 & 0 & 0
   \\
 0 & 0 & 0 & 0 & 0 & 0 & 0 & 0 & 0 & 1 & 0 & 0 & 0 & 0 & 0 & 0
   \\
 0 & 0 & 0 & 0 & 1 & 0 & 0 & 0 & 0 & 0 & 0 & 0 & 0 & 0 & 0 & 0
   \\
 0 & 0 & 0 & 0 & 0 & 1 & 0 & 0 & 0 & 0 & 0 & 0 & 0 & 0 & 0 & 0
   \\
 0 & 0 & 0 & 0 & 0 & 0 & 0 & 0 & 0 & 0 & 0 & 0 & 1 & 0 & 0 & 0
   \\
 0 & 0 & 0 & 0 & 0 & 0 & 0 & 0 & 0 & 0 & 0 & 0 & 0 & 1 & 0 & 0
   \\
 0 & 0 & 1 & 0 & 0 & 0 & 0 & 0 & 0 & 0 & 0 & 0 & 0 & 0 & 0 & 0
   \\
 0 & 0 & 0 & 1 & 0 & 0 & 0 & 0 & 0 & 0 & 0 & 0 & 0 & 0 & 0 & 0
   \\
 0 & 0 & 0 & 0 & 0 & 0 & 0 & 0 & 0 & 0 & 1 & 0 & 0 & 0 & 0 & 0
   \\
 0 & 0 & 0 & 0 & 0 & 0 & 0 & 0 & 0 & 0 & 0 & 1 & 0 & 0 & 0 & 0
   \\
 0 & 0 & 0 & 0 & 0 & 0 & 1 & 0 & 0 & 0 & 0 & 0 & 0 & 0 & 0 & 0
   \\
 0 & 0 & 0 & 0 & 0 & 0 & 0 & 1 & 0 & 0 & 0 & 0 & 0 & 0 & 0 & 0
   \\
 0 & 0 & 0 & 0 & 0 & 0 & 0 & 0 & 0 & 0 & 0 & 0 & 0 & 0 & 1 & 0
   \\
 0 & 0 & 0 & 0 & 0 & 0 & 0 & 0 & 0 & 0 & 0 & 0 & 0 & 0 & 0 & 1
\end{smallmatrix}
\right).
\end{equation}
See Appendix for more examples.
\end{example}

The operation of partial transposition is important in quantum information
theory due to the Peres-Horodecki criterion for distinguishing separable and
entangled states. In the particular case of $\C^2\otimes \C^2$ system (\ie\ two
qubits), this criterion states that the state $\rho\in\mathcal{S}(\C^2\otimes
\C^2)$ is separable if and only if $\rho^{T_1}$ is positive. 

\section{Summary}\label{sec:concluding}
We have presented a simple derivation of Schmidt decomposition for pure
states and density matrices and Kraus decomposition for quantum channels.
Using matrix reordering one can easily construct matrices corresponding to the
composition of quantum channels. In particular we have discussed partial
operations, which play a prominent role in quantum information theory.

The main advantage of the presented formulae is that they can be used directly
in computer algebra systems. Full implementation of the procedures discussed in this
report can be found in the source code of the \Mathematica\ package~\cite{qi}. 
For the sake of consistency we provide some examples of the discussed procedures in 
Appendix~\ref{app:mathematica}.

\appendix
\section{Examples in \Mathematica\label{app:mathematica}}
Below we provide some examples of the discussed procedures using \Mathematica\
computing system (see \eg~\cite{hazrat10mathematica} for an introduction to
\Mathematica). The following examples are based on the \qi\ package for
\Mathematica, which can be freely downloaded from the project home page. This
package provides the implementation of various procedures helpful during the
calculation related to quantum information processing. For the full list of
functions implemented in this package see~\cite{qi}.

After the proper installation the package can be loaded as
\begin{verbatim}
<<QI`
\end{verbatim}
After loading the package one should get some information about the used version
and release date. The examples provided in this report were tested with the
version 0.3.21 of the package.

\subsection{Matrix reorderings}
\qi\ Package provides four functions for the matrix operations described in
Section~\ref{sec:algebra}.
\begin{itemize}
\item \verb+Res+ -- reshaping operation,
\item \verb+Vec+ -- vectorization operation,
\item \verb+Unres+ -- inverse map for reshaping,
\item \verb+Unvec+ -- inverse map for vectorization.
\end{itemize}

In the above functions it is assumed that vectors can be rearranged into square
matrices. Nevertheless, it is possible to rearrange a vector into a general
$m\times n$ matrix by specifying the second argument in \verb+Unres+ and
\verb+Unvec+ functions.

In the simple case of $\M_4$ we can use the above functions as
\begin{verbatim}
mA = SymbolicMatrix[a, 4];
vA = Res[mA];
mB = SymbolicMatrix[b, 4];
vB = Vec[mB];
\end{verbatim}
Here function \verb+SymbolicMatrix[a,4]+ returns $4\times4$ matrix filled with 
elements $a_{i,j}$.

The reshuffle operation can be implemented directly using the
Definition~\ref{def:reshuffle}. Unfortunately this implementation is inefficient
as it requires the calculation of $m^2\times k^2$ matrix elements in order to
reshuffle the matrix from $\M_{mk}$.

\qi\ Package provides three methods for performing the reshuffle operation in
matrices:
\begin{itemize}
    \item \verb+Reshuffle+ -- functions based on the
    Definition~\ref{def:reshuffle} and can be used to construct the reshuffle
    matrix,
    \item \verb+ReshuffleGeneral+ -- functions
    based on the index manipulation and can be used to reshuffle matrices which
    are not necessarily square,
    \item \verb+ReshufflePermutation+ -- can be used to construct permutation matrices
    for reshuffling operation.
\end{itemize} 

For example with
\begin{alltt}
mA = SymbolicMatrix[a, 4];
mR = ReshufflePermutation[2, 2];
\end{alltt}
the following should return \verb+True+
\begin{alltt}
Unres[mR.Res[mA]] == Reshuffle2[mA, 2, 2];
\end{alltt}

Each function implementing the reshuffle operation has an equivalent function
implementing the alternative reshuffling given by
Definition~\ref{def:alt-reshuffle} (\eg\ \verb+Reshuffle+ and
\verb+Reshuffle2+).

\subsection{Schmidt decomposition}
Usually Schmidt decomposition is used in the context of vectors (\ie\
elements of $\Cplx^N$). If we define maximally entangled pure state as
\begin{verbatim}
vA = MaxEnt[4]
\end{verbatim}
we can obtain its Schmidt decomposition as
\begin{verbatim}
vAsd = SchmidtDecomposition[vA, 2, 2];
\end{verbatim}
The initial vector can be reconstructed as
\begin{alltt}
Plus @@ Table[
  vAsd[[i]][[1]](vAsd[[i]][[2]]\(\otimes\)vAsd[[i]][[3]])
  \verb+{+i, 1, 2\verb+}+
]
\end{alltt}

The meaning of the '$\otimes$' symbol is defined in the \qi\ package to provide
the required shape of the output.


To demonstrate Schmidt decomposition on the space of matrices we use the
maximally entangled mixed state on $\Mn{4}$.
\begin{verbatim}
mA = Proj[MaxEnt[4]];
\end{verbatim}
Its decomposition can be obtained as
\begin{verbatim}
mAsd = SchmidtDecomposition[mA, 2, 2];
\end{verbatim}
The initial matrix can be easily reconstructed.
\begin{alltt}
Plus @@ Table[
  mAsd[[i]][[1]](mAsd[[i]][[2]]\(\otimes\)mAsd[[i]][[3]]), 
  \verb+{+i, 1, 4\verb+}+
]
\end{alltt}

Note that \verb+SchmidtDecomposition+ function works for vectors as well as for
matrices. However, it is possible to use \verb+VectorSchmidtDecomposition+ and
\verb+OperatorSchmidtDecomposition+ for an appropriate input instead.

\subsection{Quantum channels}
\qi\ Package defines quantum channels using \emph{pure functions} mechanism. For
example, the transposition map can be implemented as
\begin{verbatim}
TransposeChannel =
 IdentityMatrix[#1].Transpose[#2]&;
\end{verbatim}
and its matrix representation can be obtained as
\begin{verbatim}
mT = ChannelToMatrix[TransposeChannel[#]&, 4];
\end{verbatim}
for a map acting on $\M_4$. 

Similar construction for the swap operation reads
\begin{verbatim}
SwapChannel = Swap[#1].(#2).Swap[#1] &;
\end{verbatim}
and this function requires information about the system dimension. Here we have
used the \verb+SWAP+ gate predefined in the package. For example the \SWAP\ operation
on two qubits is defined as
\begin{verbatim}
cS4 = SwapChannel[4,#];
\end{verbatim} 

Again, one can obtain a matrix representation of this channel as
\begin{verbatim}
mS = ChannelToMatrix[cS4, 4];
\end{verbatim}
Alternatively, the same result can be obtained using 
\begin{verbatim}
mS = Superoperator[cS4, 4];
\end{verbatim}

\subsubsection{Spontaneous emission channel for qutrits}
Following~\cite{checinska07eparability} (see also~\cite{checinska06noisy},
\qi\ package provides a definition of a spontaneous emission channel for a three-level
 system (qutrit).
\begin{verbatim}
seK = QutritSpontaneousEmissionKraus[A1,A2,t];
\end{verbatim}
Here \texttt{A1} and \texttt{A2} are Einstein coefficients.

This channel was used \eg\ in~\cite{gawron08noise} to investigate the behaviour
of quantum games under decoherence.

The superoperator corresponding to the above channel can be obtained as
\begin{verbatim}
seS = Superoperator[seK];
\end{verbatim}

Note that the \texttt{Superoperator} function has two forms and can be used to
obtain matrix representation of the channel either from the list of Kraus
operators or from the pure function.

\subsubsection{Partial operations}\label{sec:qi-partial-ops}
The notion of partial operation is very common in quantum information theory and
the presented package allows to construct and analyse such operations in a very
straightforward manner.

Let us consider an operation $\Psi$ on $n$-dimensional system defined as a
pure function \texttt{fPsi}. In order to obtain the operation $\Psi\otimes\Id$
acting on $n\times m$-dimensional system one needs to
\begin{itemize}
  \item construct the matrix representation of the map $\Psi$:
  \begin{verbatim}sPsi = Superoperator[fPsi,n]\end{verbatim}
  \item construct the reshuffle matrix in order to transform the obtained matrix 
  to a new base matrix of the appropriate size:
  \begin{alltt} mR = ReshufflePermutation[n n, m m]\end{alltt}
  \item use the matrix \texttt{sPsi} according to the
  Def.~\ref{def:composite-channel}, using the $\Id$ channel on the second
  subsystem:
  \begin{alltt}extPsi = mR.(sPsi\(\otimes\)IdentityMatrix[m\^{}2]).mR\end{alltt}
\end{itemize}
The extension of the operation constructed in the above procedure acts on the 
$n\times m$-dimensional states $\rho$ as
\begin{alltt}
Unres[extPsi.Res[\(\rho\)]]
\end{alltt}

Note that this procedure requires to construct a matrix of dimension which grows
like $\mathcal{O}(n^4)$ and can be slow for larger systems.

The simplest case of such construction is the partial transposition on
$n^2$-dimensional space. The matrix representation of the transposition
operation of dimension $n$, is a \SWAP\ operation on $n^2$-dimensional system.
The matrix representation of the partial transposition can be obtained as
\begin{alltt}
mR = ReshufflePermutation[n n, n n];
tA = mR.(Swap[n n]\(\otimes\)IdentityMatrix[n n]).mR
\end{alltt}

Similar procedure is implemented in \qi\ as \texttt{PartialTraceA} function.
However, as this implementation is not very efficient, \qi\ provides an
alternative version of this operation as a \texttt{PartialTraceGeneral}
function, which operates on indices. For example, for a given matrix
\begin{alltt}
mA = SymbolicHermitianMatrix[a, 4]\end{alltt}
the results of
\begin{alltt}
PartialTraceA[mA, 2, 2]\end{alltt}
and
\begin{alltt}
PartialTraceGeneral[mA, {2, 2}, 1]\end{alltt}
are identical.

The matrix representation of the partial transposition with respect to the first
subsystem, given in~Eq.~(\ref{eqn:ptrace-matrix}), can be obtained as
\begin{alltt}
Superoperator[PartialTransposeA[#, 2, 2] &, 4]
\end{alltt}
Here \verb+PartialTransposeA[#,2,2]+ is a map which is not positive.

\begin{acknowledgments}
This report was initiated by the discussion with Z.~Pucha\l{}a and P.~Gawron in
Brno in November 2008. It is a pleasure to thank J. Bouda for hosting us at the
Faculty of Informatics of the Masaryk University in Brno.

I~would also like to acknowledge the financial support by the Polish Ministry of
Science and Higher Education under the grant number N N519 442339.
\end{acknowledgments}

\bibliography{kraus_schmidt}
\bibliographystyle{apsrev4-1}

\end{document}